# Improving Web Page Readability by Plain Language

Walayat Hussain[1], Osama Sohaib[2] and Arif Ali[3]

[1] Department of Computer Science, Balochistan University of I.T. Engineering and Management Science,
Quetta, Pakistan

Department of CS & IT, University of Balochistan
Quetta, Pakistan

[3] Department of Information Technology, Balochistan University of I.T. Engineering and Management Science
Quetta, Pakistan

**Abstract**
In today's world anybody who wants to access any information the first choice is to use the web because it is the only source to provide easy and instant access to information. However web readers face many hurdles from web which includes load of web pages, text size, finding related information, spelling and grammar etc. However understanding of web pages written in English language creates great problems for non native readers who have basic knowledge of English. In this paper, we propose a plain language for a local language (Urdu) using English alphabets for web pages in Pakistan. For this purpose we developed two websites, one with a normal English fonts and other in a local language text scheme using English alphabets. We also conducted a questionnaire from 40 different users with a different level of English language fluency in Pakistan to gain the evidence of the practicality of our approach. The result shows that the proposed plain language text scheme using English alphabets improved the reading comprehension for non native English speakers in Pakistan.

## 1. Introduction

In today's life World Wide Web has been considered as an instant and easy source to get any information. No doubt the World Wide Web contributes greatly in creation of an increasing global information database. Using web site anyone can promote its ideas business very easily. Almost all households have access to the internet and can reach the whole world in just few clicks. Beside the uses of internet there are lots of problems associated with the web like lost of web pages, web pages load, and text too small or too large, bad spelling or grammar, finding related information, appropriate guidance for users to relevant information and when get some information then the understanding of these information [15]. Among all these problems one of the major problem is the understanding of the text and materials written in website. In today's web primary language for the websites are English language due to which non native readers whose primary language is not English faces great problems due to their limited knowledge about unfamiliar vocabulary, the grammar, composition and structure of sentences, self explanatory graphs, and use of abbreviations or intimidating content display [1], which creates lots of problems in web readability. Web readability can be defined as "*a combination of reading comprehension, reading speed and user satisfaction in terms of reading comprehension, dictionary, thesaurus and existing online tools and browser add-ones*". Readability may also be defined as "*how easily a person can read and understand any written materials*". Website readability is an indicator of overall difficulty level of a website [1][2].

Many researchers have discussed web readability issues and proposed various ideas to enhance web readability [1][2][3][4], this study also discuss web readability by addressing following research question.

- How to enhance web readability for users whose first language is not an English language?
- Which approach has been adopted to cope up readability issue while using English language and Local languages?

In this paper, we propose a plain language text using English alphabets for web pages. For this purpose we developed two websites, one with a normal English alphabet and other in a local language (Urdu) text scheme using English alphabets with a questionnaire to evaluate our proposal. We conducted a study on 40 different users

316

with a different level of English language fluency in Pakistan.

The work is organized as follows. Section 2 gives the overview and background of the internet users in Pakistan. Section 3 present some related studies. Section 4 provides the approach of our study. Section 5 and 6 shows interpretations. Finally the study is concluded and leaves some an open issue.

## 2. Background and Motivation

Plain Language is a writing approach that is effective to understand information easily the first time reader's reads it [14].

Pakistan is a multilingual country, it has two official languages: English and Urdu. Urdu is also the national language. Additionally, Pakistan has four major provincial languages Punjabi, Pashto, Sindhi, and Balochi, as well as two major regional languages: Saraiki and Kashmiri [10].

Table 1: Pakistani languages

| Languages | Percentage of speakers |
|---|---|
| Punjabi | 44.15 |
| Pashto | 15.42 |
| Sindhi | 14.10 |
| Siraiki | 10.53 |
| Urdu | 7.57 |
| Balochi | 3.57 |
| Other | 4.66 |

Internet access has been available in Pakistan since 1990s. The country has been following an aggressive IT policy, aimed at enhancing Pakistan's drive for economic modernization and creating an exportable software industry. There is no doubt that has been helping increase the popularity of the Internet. Table 2 shows the number of users within a country that access the Internet [11].

Table 2: Internet users in Pakistan

| Year | Internet users | Rank | Percent Change | Date of Information |
|---|---|---|---|---|
| 2003 | 1,200,000 | 47 |  | 2000 |
| 2004 | 1,500,000 | 48 | 25.00 % | 2002 |
| 2005 | 1,500,000 | 49 | 0.00 % | 2002 |
| 2006 | 10,500,000 | 23 | 600.00 % | 2005 |
| 2007 | 10,500,000 | 24 | 0.00 % | 2005 |
| 2008 | 17,500,000 | 17 | 66.67 % | 2007 |
| 2009 | 17,500,000 | 17 | 0.00 % | 2007 |
| 2010 | 18,500,000 | 20 | 5.71 % | 2008 |

Where English is also an official language but it is not the most spoken language of Pakistan. Because English is so widely spoken, it has often been referred to as a world language [12]. That is why English is taught as foreign language in Pakistan, but still the percentage of English fluency is low among the people in Pakistan. The Literacy rate of Pakistan is (56%). Sindh (58%) and Punjab (58%) are equally more literate as compared to NWFP (50%) and Balochistan (49%) provinces. The percentage of English speakers in the country is only 10.9% [13]. So the users when try to read the information on the web in English, they suffer with web readability. We try to solve this problem by using English alphabets written in local language (Urdu). Although the Google translation of the web pages from English to Urdu (national language of Pakistan) is available, but the main problem with that, it translate the sentence word by word, which does not make the Urdu sentence understandable

Major headings are to be column centered in a bold font without underline. They need be numbered. "2. Headings and Footnotes" at the top of this paragraph is a major heading.

## 3. Related Work

Web becomes more complex with the fast growth of information distributed through web pages especially that use a fashion-driven graphical design but readability of WebPages is not taken into consideration. The readability is an important criterion for measuring the web accessibility especially non-native readers encounter even more problems.

Readability crucial presentation attributes that web summarization algorithms consider while generating a query based web summary. Text on the web of a suitable level of difficulty for rapid retrieval but appropriate techniques needs to be work out for locating it. Readability measurement is widely used in educational field to assist instructors to prepare appropriate materials for students. However, traditional readability formulas are not fit to attract much attention from both the educational and commercial fields [1][2][5][6][7][8][9].

Miller and Hsiang Yu [1] propose a new transformation method, Jenga Format, to enhance web page readability. A user study on 30 Asian users with moderate English fluency is conducted and the results show that the proposed transformation method improved reading comprehension without negatively affecting reading speed. The authors have solved the problems of distraction elimination and content transformation. They have found two important factors, sentence separation and sentence spacing, affecting the reading.

Pang Lau and King [2] propose a bilingual readability assessment scheme for web site in English and Chinese languages. The Experimental results show that, for page readability apart from just indicating difficulty, the estimated score acts as a good heuristic to figure out pages with low textual content such as index and multimedia pages.

Gradišar et al [5] identifies the existence of factors that influence reading experience, the authors examined the readability of combination of 30 different text colors that are presented on the CRT display by measurement of speed of reading through Chapman-Cook Speed. The results show that there are no statistically major differences in readability between 30 color combinations but they have prove an existence of at least five factors, which simultaneously and differently affect readability of a colored text.

Uitdenbogerd [6] experimented and compare the range of difficulty of the text web that is found in traditional hard-copy texts for English as Second Language (ESL) learners using standard readability measures. The results suggest that an on-line text retrieval engine based on readability can be of use to language learners because of the ESL text readability range fall within the range for web text.

Xing et al [7] demonstrates a novel approach, in order to increase the accuracy of readability for measuring English readability applying techniques from natural language processing and information theory. The authors have found by applying the concept of entropy in information theory that the readability differences are not caused by the text itself but by the information gap between text and reader.

Gottron and Martin [8], describes the modern content extraction algorithms that help to estimate accurately the readability of a web document prior to index calculation. The authors observed the SMOG and the FRE index to be far more accurate in combination with CE in comparison to calculating them on the full document.

Kanungo and Orr [9] propose a machine learning methodology that first models the readability of abstracts using training data with human judgments, and then predicts the readability scores for previously unseen documents using gradient boosted decision trees. The performance of the model goes beyond that of other kinds of readability metrics such as Collins-Thompson-Callan, Fog or Flesch-Kincaid. The model can also be used in the automatic summarization algorithm to generate summaries that are more readable

## 4. The Approach

In order to understand the effect of content transformation and to analyze the difference and compare the readability between English language and local language written in English alphabet (plain language) we developed two websites with four web pages each and conducted a formal user study to investigate the effectiveness of both contents from end users point of view [Table 3].

Table 3: web pages information

|  | Page 1 | Page 2 | Page 3 | Page 4 |
|---|---|---|---|---|
| Website 1 | Passage A (English) | Questionnaire | Passage B (Translated) | Questionnaire |
| Website 2 | Page 1 | Page 2 | Page 3 | Page 4 |
|  | Passage A (Translated) | Questionnaire | Passage B (English) | Questionnaire |

Websites contents are from standard IELTS test. In first website first page has small passages written in English language, and second page had another same size passage as first but this was translated into national language (Urdu) using English alphabets. In second website the first page had the same content as the first page of first website but this was translated passage, and the second page had the English version of the second page of first website. At the end of each passage there is a questionnaire which included few MCQ's related to that passage.

The website first register the users, then a reader started the test by reading a first passage after completing questions, at the end of a each test a reader moved to the next test with another passage translated into national language using English alphabet. For each test there were timers which take the duration of time spent on each test. After completing both tests there is another page for the feed back in order to comments about both tests.

## 5. Results

There are 40 users selected for this test: 12 females and 28 males. All the users are from Pakistan whose first language are not English. We have categorized our users into three categories [Table 4]:
- Undergraduate Students.
- Professionals having good knowledge of English language
- Other Workers having basic knowledge of English language

Table 4: user's categories

| Category | Education | No. of Users |
|---|---|---|
| Professionals | Masters / Bachelors | 10 |
| Students | Undergraduate | 18 |
| Workers | Secondary School | 12 |

At the end of each passage there are nine questions related to that passage. The result of each group is shown in Table 5 and Figure 1. The time taken in reading both passages of English and a plain language (Local language-Urdu) written in English alphabets is shown in Figure 2.

Table 5: Correct answers attempt by users with time taken

| No. of correct answers attempted by Students | | | | |
|---|---|---|---|---|
| No. of | English | Time/User | Translated | Time/User |

| cases | Pages | | pages | |
|---|---|---|---|---|
| 18 | 50% | 9min | 84% | 11min |
| **No. of correct answers attempted by Workers, Lower literate user** | | | | |
| No. of cases | English | Time/User | Translated | Time/User |
| 12 | 9% | 40min | 86% | 19min |
| **No. of correct answers attempted by Professionals** | | | | |
| No. of cases | English | Time/User | Translated | Time/User |
| 10 | 88% | 8min | 90% | 10min |

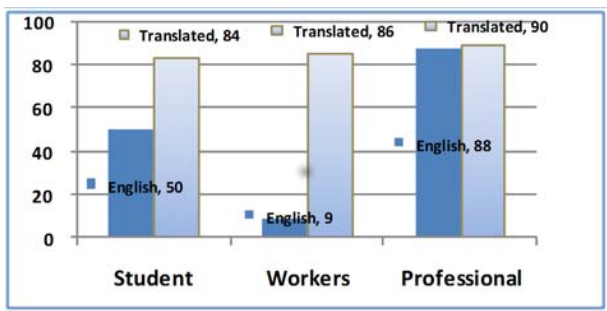

Figure 1: Percentage of correct answers of both tests

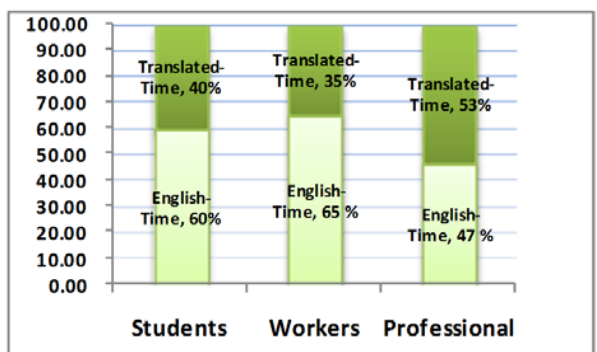

Figure 2: Percentage of Time taken for both test (English passages and Translated passages)

## 5. Findings

Based on the results of the study, we can say that:
1. The transformation of the text content enhances web readability for non native user i.e. whose first language is not English.
2. The translated version of an English text gives better result and the percentage of correct answers is more than English passage text.
3. The ratio of correct answers for translated version is very high in Worker, lower-literate user and in undergraduate student's category because they have only basic and moderate knowledge of English. In Student category there is slight difference while for professionals who have good understanding of English and have high education is almost equal.
4. As in Pakistan the level for higher education is less than the population having basic education, so the translated version of English text is more readable than English text.

There are many interesting comments from users which they have given at the end of test. Like most of the people are not very much familiar with the translated version that's why it takes longer time for them and they preferred for English version of a text although they made more mistakes while solving the questions of English passage compared to translate version. Many users recommend the translated version of a passage because of easily understandable.

## 5. Conclusion and Future Work

In order to analyze the difference and to compare the readability between English language and a pain language (local language -Urdu) using English alphabets we develop two sites having few pages. We take two passages from the standard IELTS reading passage. At the end of each passage there is a questionnaire for investigating the effect of our approach.

We have observed that by changing the contents of web pages into local language by using English alphabets we get better result. The level of understanding content is very much high for all those who have basic knowledge of English. By using this approach we can enhance web readability to all those non native English speaker countries whose local language text is incompatible with the web pages.

We plan to extend our study in other Asian countries for the non native readers to investigate the usefulness of our approach to web readability.

**Walayat Hussain** received the BS (Software Development) Hons degree from Hamdard University Karachi, Post Graduation in Computer Science from AIT (Asian Institute of Technology) - Bangkok and MS (Computer Science) degree from BUITEMS Quetta, Pakistan in 2004, 2008 and 2009 respectively. He is currently working as an Assistant Professor in Department of CS in Balochistan University of IT, Engineering and Management Sciences (BUITEMS), Quetta Pakistan.

**Osama Sohaib** received the BS (Software Development) Hons degree from Hamdard University Karachi, Post Graduation in Information Management from AIT (Asian Institute of Technology) - Bangkok and MS (Software Engineering) degree from PAF-KIET (Karachi Institute of Economics and Technology), Karachi Pakistan, in 2005, 2008 and 2010 respectively. He is currently working as a Lecturer in Department of CS & IT in University of Balochistan, Quetta Pakistan.

**Arif Ali** received the BSc (Business Information Technology) Hons degree from University of Salford UK in 2006 and MSc Business Information Systems from University of Bolton UK in 2008. He is currently working as Lecturer at Department of IT at Balochistan University of IT, Engineering and Management Sciences (BUITEMS), Quetta Pakistan since November 2009.